\documentclass[referee]{raa}            
\usepackage{graphicx,times}             
\usepackage{natbib}
\usepackage{amssymb,amsmath}
\usepackage{mathrsfs}
\bibpunct{(}{)}{;}{a}{}{,}
\usepackage[a4paper,left=2.5cm,right=2.5cm,top=2.5cm,bottom=2.5cm]{geometry}
\usepackage[a4paper=true,
pagebackref=true]{hyperref}
\hypersetup{colorlinks = true, linkcolor = green, anchorcolor = red, citecolor = blue, filecolor = red, pagecolor = red, urlcolor = red}

\begin{document}
\title{Effect of Anisotropic Collisions on Solar Scattering Polarization
}
%
\volnopage{Vol.0 (20xx) No.0, 000--000}      
\setcounter{page}{1}          

\author{Saleh Qutub \inst{1}   \and   Moncef Derouich \inst{1,2}   \and    Badruddin Zaheer Ahmad \inst{1}}
%
 \institute{ Astronomy and Space Science Department, Faculty of Science, King Abdulaziz University, Jeddah 21589, Saudi Arabia 
\and
Sousse University, ESSTHS, Lamine Abbassi street, 4011 H. Sousse, Tunisia \\
{\small Received~~20xx month day; accepted~~20xx~~month day}}
\abstract{Scattering of anisotropic radiation by atoms, ions   or  molecules is sufficient to generate  linear polarization observable in stars and planets atmospheres, circumstellar environments, 
 and in particular in the Sun's atmosphere. This kind of polarization is called scattering polarization (SP)
  or second solar spectrum (SSS) if it is formed near the limb of the solar photosphere. 
 Generation of linear 
SP can  typically  be reached more easily than circular SP. Interestingly, the later is often absent in observations and theories.
 Intrigued by this, we propose to demonstrate how circular SP can be created by anisotropic collisions if a magnetic field 
 is present. 
We also demonstrate how anisotropic collisions can result in the creation of circular SP if the radiation field is anisotropic.
 We show that under certain conditions, linear SP creation is accompanied by   the emergence of  circular SP  which can be  useful for diagnostics  of  
solar and astrophysical plasmas. We treat an example and calculate  the density matrix elements of tensorial order $k\!=\!1$ which are directly associated with the presence of circular SP.  
This work should encourage
theoretical and observational research to be increasingly oriented towards
circular SP profiles in addition to linear  SP in order to improve  our analysis tools of astrophysical and solar observations.
}

\authorrunning{ Qutub et al.}  
\titlerunning{Effect of   Anisotropic Collisions }  

\maketitle

\section{Statement of the problem}	\label{sect:intro} 

Symmetry-breaking processes, such as (de-)excitations by anisotropic light  or anisotropic collisions,  could generate the so-called scattering  polarization (SP) of the  emitted  light.
An atom
is   said to be  polarized by scattering if 
  the scattering processes result in uneven population of its Zeeman sublevels and thereby the appearance of coherences between them.
This
is what is referred  to  as atomic polarization  (e.g. Sahal-Br\'echot 1977, Trujillo Bueno 2002, Sect. 3.6 of Landi Degl'Innocenti 2004).  For an in-depth understanding of astrophysical/solar plasma,  the polarization  properties
of light emitted by atoms/ions/molecules  must be 
carefully studied from an observational and theoretical  points of view.  In this context, newer theoretical techniques and modern instruments allowing observation and interpretation of small polarization signals are needed.

The effect  of collisions on  atomic states  $| \alpha J \rangle$,  and therefore on 
 the SP, 
 can be described by   the 
polarization transfer  and relaxation  rates;   here  $J$  denotes  the total angular momentum  and $\alpha$ represents the other  quantum numbers associated   with  the atomic state. 
  For the study of polarization of spectral lines it is more convenient to use the density matrix formalism expressed on the basis of irreducible tensorial operators, $T^k_q$. In this framework, the density 
 matrix elements $\rho_{q}^{k}  ( \alpha J)$  give the average state of the  polarized atom which
emits the   polarized light   (see e.g. Sects. 3.6 and 3.7 of Landi Degl'Innocenti  \& Landolfi 2004). 
Here  $k$ is the tensorial order and $q$ is the coherence between the Zeeman sublevels, where $0  \le  k \le 2J$ and $-k \le q \le k$.
The
element $\rho^{k=0}_{q=0} (\alpha J$) 
is related to the population of the $J$-level whereas 
 elements with $k \!\ge\! 1$ characterize the polarization state of the atom and consequently of the emitted radiation. In particular, the circular SP represents the observational signature of the  {\it orientation} of atomic levels and is quantified  by the density matrix elements  with odd rank, $\rho_{q}^{k=1} (\alpha J$), $\rho_{q}^{k=3} (\alpha J$), etc., while linear SP is associated with the atomic level {\it alignment} which is characterized by even tensorial order density matrix elements, $\rho_{q}^{k=2} (\alpha J$), $\rho_{q}^{k=4} (\alpha J$), etc.

In the solar context, observations with the TH\'EMIS  telescope (Spain) and with  the Advanced Solar Polarimeter (USA)
    by
L\'opez Ariste et al. (2005)  have revealed   the existence of  unexpected circular SP (symmetric $V$-Stokes)
  of the
  H$\alpha $ line which cannot
be attributed to the Zeeman effect. 
On the contrary, by using ZIMPOL telescope, Ramelli et al. (2005) observed
$V$ profiles   showing  
antisymmetric shape  typically due to the Zeeman effect.
Let us
recall that
symmetric $V$-Stokes profiles are related to the circular SP and hence to  the orientation of the atomic level (i.e. $\rho_{q}^{k} (\alpha J)$ with odd $k$),
while
anti-symmetric $V$-Stokes are known to be 
due to the Zeeman effect. 
In the light of these contradictory observations, theoretical  interpretation 
seems to be necessary. Casini
\& Manso Sainz (2006) proposed that the observation of symmetric $V$-Stokes
  could be due to the effect of an electric field. Derouich (2007) proposed a scenario based on impact circular polarization by anisotropic collisions. 
In addition,  linear-to-circular  SP transfer processes have been highlighted theoretically by Manabe et al. (1979) and measured   experimentally  by the same authors  in 1981 (Manabe   et al.   1981).  Similar processes have  been reported also by  Petrashen'
  et al. (1993), which also contains extensive references.

    It is well known that isotropic collisions can only result in the decrease of atomic polarization  (e.g. Derouich et al. 2003).
However, 
anisotropic collisions can create or increase
  the polarization of   $  | \alpha  J  \rangle $ levels.   The variation  of the atomic polarization may be due to   transitions between Zeeman sublevels of the same electronic level $  | \alpha  J  \rangle $  and/or between two different electronic levels.
 This can roughly be interpreted as the transfer of anisotropy from the relative velocity distribution of the colliding partners to the population of the Zeeman sublevels of the electronic sublevels involved in the transitions (e.g. D'yakonov \& Perel 1978).

\begin{figure} 
\centering{\includegraphics[width=9.0cm]{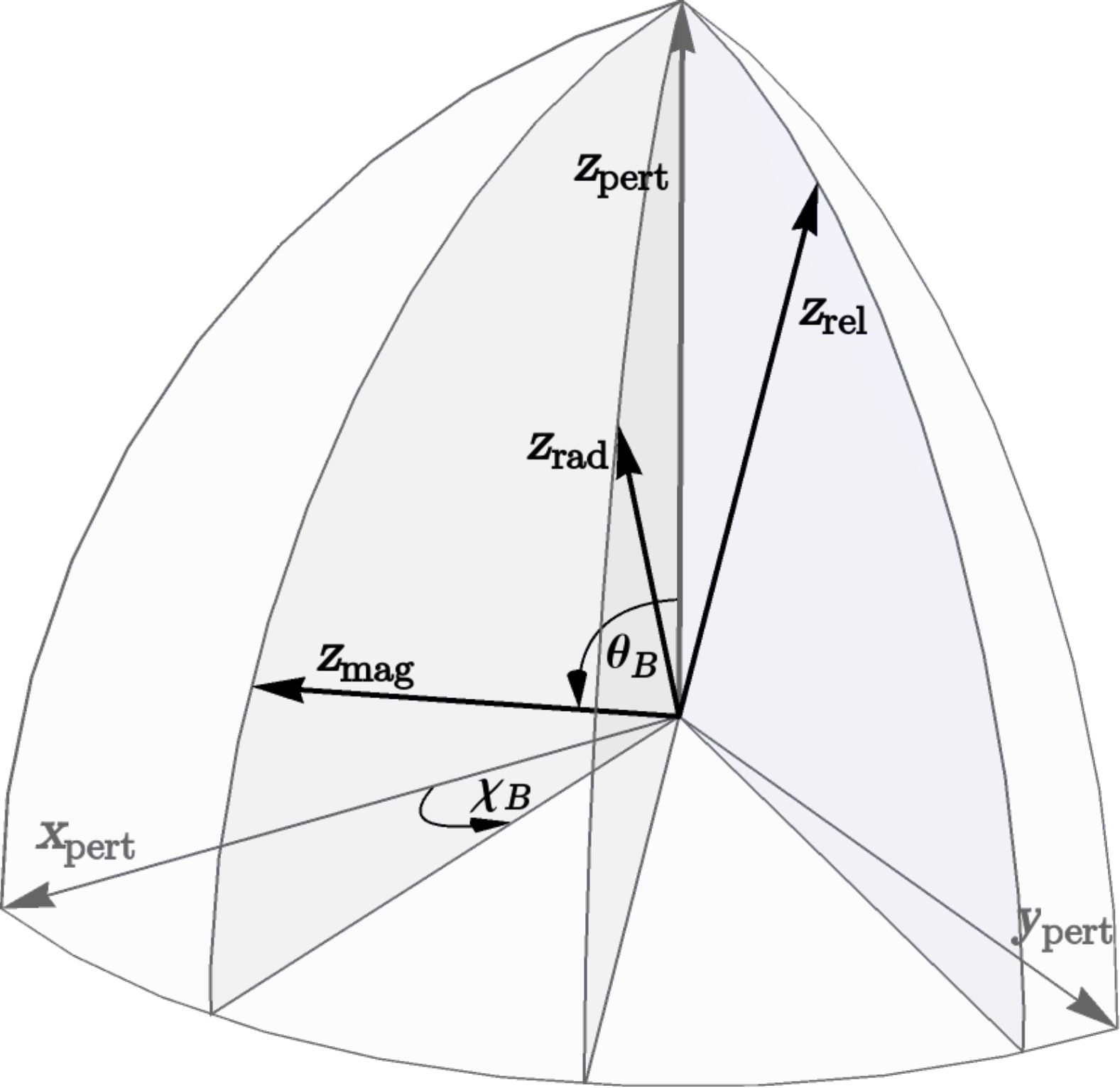}}
\caption{Geometrical configuration of the different axes drawn in the reference frame $\Sigma$. For a given binary collision, the relative velocity of the colliding partners, $v_{\rm rel}$, points in the direction of the emitter-perturber axis, $z_{\rm rel}$.}
\label{fig:config}
\end{figure}

Now
consider an   ensemble of atoms    illuminated by  unpolarized  light
  having  cylindrical  symmetry around an axis $ z_{\rm rad} $.  The  atoms also undergo  anisotropic collisions  with  beams of perturbers having axial symmetry around an axis $ z_{\rm pert} $. 
Furthermore, in a magnetized plasma like the Sun, Hanle effect of a magnetic field is an important ingredient in the modeling of the polarization state (e.g. Hanle 1924, Sect. 10.3 of Landi Degl'Innocenti \& Landolfi 2004, Derouich et al. 2006, del Pino Aleman et al. 2018). Let us  therefore  consider a general case of a magnetic field oriented   along  an axis $ z_{\rm mag} $ which is neither parallel to $ z_{\rm rad} $ nor parallel to  $ z_{\rm pert} $.  The geometrical configuration of the different axes is depicted in Figure~\ref{fig:config}.
Our aim in this work is to demonstrate that,   under  these conditions,
 mixing between  even and odd tensorial orders
is allowed and can be highlighted theoretically and observationally by obtaining non zero   circular SP (i.e. symmetric $V$-Stokes signals).

\section{Theoretical Considerations  } \label{sec:theory} 
In order  to solve  the   statistical equilibrium equations (SEE) for the  atomic   levels  $| \alpha_{i} J_{\alpha_{i}} \rangle$  of the emitting atom described by the elements $ \rho^k_q(\alpha_{i} J_{\alpha_{i}})$, we place ourselves in a  reference frame, $\Sigma$, centered on the atom and having its $z$-axis in the $ z_{\rm pert} $ direction. The frame $\Sigma$ is  obtained by a rotation ${\rm R_B} \equiv {\rm R}(\text{-}\gamma_{\rm B},\text{-}\theta_{\rm B},\text{-}\chi_{\rm B})$ of the magnetic reference defined by the  $z_{\rm mag}$-axis (see Figure~\ref{fig:config}).\footnote{Since the magnetic kernel ${\mathcal K}^k_{qq'} ({\rm R_B})$ is independent of  the Euler angle $\gamma_{\rm B}$, it can be arbitrarily set to zero, $\gamma_{\rm B} \!=\!0$ (see e.g. page 548 of Landi Degl'Innocenti \& Landolfi 2004).} 
The collisional cross sections are usually obtained in the collision frame having its $z-$axis
 joining the perturber and the perturbed atom ($z_{\rm rel}$ in Figure~\ref{fig:config}), and then rotated to the frame $\Sigma$ where the average over relative velocity distribution is performed. In fact in the latter frame, the symmetry properties of the collisional rates are manifest which simplifies the solution of SEE. The radiative contributions to the SEE are also rotated to the frame $\Sigma$. 
It can be proved that 
  the  expressions  for  relaxation and transfer radiative rates are formally invariant under  rotation (see e.g. pages 330-331 of Landi Degl'Innocenti \& Landolfi 2004).  

In the basis of irreducible tensorial operators
and
in the reference $\Sigma$,
the  time variation of the elements $ \rho^k_q(\alpha_{i} J_{\alpha_{i}})$ 
 can be written as~(see e.g. pages 284-285 of Landi Degl'Innocenti \& Landolfi 2004, Manabe et al. 1979):
\begin{align} \label{eq:SEE1}
\!\!\!
\frac{\rm d}{{\rm d} t} \rho^k_q(\alpha_{i} J_{\alpha_{i}})
\!=&\!
- {\rm i} \,  \omega_{{\rm L},\alpha_{i} J_{\alpha_{i}}}  \, g_{\alpha_{i} J_{\alpha_{i}}}  \!\! \sum_{q'} \! 
{\mathcal K}^k_{qq'} ({\rm R_B}) \, \rho^k_{q'}(\alpha_{i} J_{\alpha_{i}})
- \! \sum_{k'q'} \! \left[ r_{\rm A}(\alpha_{i} J_{\alpha_{i}} kq k'q') + r_{\rm E}(\alpha_{i} J_{\alpha_{i}} k q k'q')  \right]
\rho^{k'}_{q'}(\alpha_{i} J_{\alpha_{i}}) 
\nonumber \\ \!\!\!&\!\!\!
+\!
\sum_{ \substack{j k' q' \\ j<i} } \!
t_{\! \rm A}(\alpha_{i} J_{\alpha_{i}} k q, \alpha_{j} J_{\alpha_{j}} k' q')
\,  \rho^{k'}_{q'}(\alpha_{j} J_{\alpha_{j}})
+
\!  \sum_{\substack{j k' q' \\ j>i} } \! 
t_{\! \rm E}(\alpha_{i} J_{\alpha_{i}} k q, \alpha_{j} J_{\alpha_{j}} k' q') \,  \rho^{k'}_{q'} (\alpha_{j} J_{\alpha_{j}})
\nonumber \\ \!\!\!&\!\!\!
+\!
\sum_{ j k' q' } \!
\mathcal{T}^{k k'}_{q q'} (\alpha_{i} J_{\alpha_{i}} \!\leftarrow\!  \alpha_{j} J_{\alpha_{j}}) \, \rho^{k'}_{q'} (\alpha_{j} J_{\alpha_{j}}) 
-\!\! \sum_{ j k' q' } \!  \mathcal{R}^{k k'}_{q q'}(\alpha_{i} J_{\alpha_{i}} \!\to\!  \alpha_{j} J_{\alpha_{j}})  
\rho^{k'}_{q'}(\alpha_{i} J_{\alpha_{i}})    \, .
\end{align} 
For simplicity, we have ignored stimulated emissions since they are   negligible in natural plasma such as the solar atmosphere. The   term $- {\rm i} \,  \omega_{{\rm L},\alpha_{i} J_{\alpha_{i}}}  \, g_{\alpha_{i} J_{\alpha_{i}}}  \!\! \sum_{q'} \! 
{\mathcal K}^k_{qq'} ({\rm R_B}) \, \rho^k_{q'}(\alpha_{i} J_{\alpha_{i}})$ gives   Hanle   effect of magnetic field in the $\Sigma$ reference, where $\omega_{{\rm L},\alpha_{i} J_{\alpha_{i}}} \!=\! 2 \pi \nu_{{\rm L},\alpha_{i} J_{\alpha_{i}}}  $ denotes the Larmor angular frequency and $g_{\alpha_{i} J_{\alpha_{i}}}$ denotes the Land\'{e} g-factor.
the expression for the magnetic kernel ${\mathcal K}^k_{qq'} ({\rm R_B})$ can be found, for example, on page 548 of Landi Degl'Innocenti \& Landolfi (2004). 
$r_{\rm E}$ and $r_{\rm A}$   respectively  denote the relaxation rates due to spontaneous emission and absorption, 
while  
$t_{\! \rm E}$ and $t_{\! \rm A}$  respectively  denote the transfer rates due to spontaneous emission and absorption.
Expressions
for these radiative rates can  be found in the literature (see e.g.  Bommier \&  Sahal-Br\'echot 1978, pages 287-288 of Landi Degl'Innocenti \& Landolfi 2004).

We note here that
due to mixing under rotation, 
the coherences in the radiation field tensor, ${\rm J}^{k_{\rm r}}_{q_{\rm r}}$ responsible for the radiative rates $r_{\rm A}$ and  $t_{\rm A}$, are present in the frame, $\Sigma$, despite being non-existent in the radiation frame.  
$\mathcal{T}^{k k'}_{q q'}(\alpha_{i} J_{\alpha_{i}} \!\leftarrow\!  \alpha_{j} J_{\alpha_{j}})$
and
$\mathcal{R}^{k k'}_{q q'}(\alpha_{i} J_{\alpha_{i}} \!\to\!  \alpha_{j} J_{\alpha_{j}}) $ denote the collisional transfer and relaxation rates, respectively.
The quantity $\mathcal{T}_q^{kk'}(\alpha_{i} J_{\alpha_{i}} \!\leftarrow\!  \alpha_{j} J_{\alpha_{j}})$ represents the gain due to collisional transitions from other levels ($j\!\neq\!i$) and sublevels of the same level ($j \!=\! i$) in contrast to $\mathcal{R}_q^{kk'}(\alpha_{i} J_{\alpha_{i}} \!\to\!  \alpha_{j} J_{\alpha_{j}})$ which represents the relaxation (loss) due to collisional transitions to other levels ($j\!\neq\!i$) and sublevels of the same level ($j \!=\! i$). In the dyadic    basis, 
for collisional transition taking place within the same level ($j \!=\! i$), 
$\mathcal{R}_q^{kk'}$ is associated to the term
$\sum_{ M' \ne M} \mathcal{R}_{\rm elastic}(\alpha_{j} j  M  \!\rightarrow\! \alpha_{j} j  M' ) \times   \rho(\alpha_{j} j  M  ) $ 
and 
$\mathcal{T}_q^{kk'}$ contains the term $  \sum_{ M'   \ne M} \mathcal{T}_{\rm elastic}(\alpha_{J} J  M  \leftarrow \alpha_{J} J  M' )  \times    \rho(\alpha_{J} J  M' )$
(see e.g. Derouich et al.  2003). 
The rates $\mathcal{T}_{\rm elastic}$ and $\mathcal{R}_{\rm elastic}$ are not equal since they are related to two different transitions.
Therefore, the quantities $\mathcal{T}_q^{kk'}(\alpha_{i} J_{\alpha_{i}} \!\leftarrow\!  \alpha_{i} J_{\alpha_{i}})$ and $\mathcal{R}^{k k'}_{q q'}(\alpha_{i} J_{\alpha_{i}} \!\to\!  \alpha_{i} J_{\alpha_{i}}) $ are in general different in value. It is to be noticed that  even if they are  equal our solutions of the SEE remain valid and, as it can be verified in the next section, the orientation emergence will be clearly possible.
The expression for the collisional rates in the case of axial symmetry   around the axis $ z_{\rm pert} $
can be found, for example, in Derouich (2007) and Manabe et al. (1979).

Axial symmetry of collisions combined with the Hermiticity of the density matrix dictate that~(e.g. Omont 1977 and Manabe et al. 1979):
\begin{align} \label{eq:ColSym2}
\mathcal{T}^{k k'}_{qq'} &= 0 \,\,\, \text{ for } q \neq q' \,\,\, \text{ i.e.} \,\,\,
\mathcal{T}^{k k'}_{qq'} = \delta_{qq'}  \mathcal{T}^{k k'}_{qq}=   \mathcal{T}^{k k'}_{q}
 \nonumber \\
\mathcal{T}^{k k'}_{\text{-}q} &= (\text{-}1)^{k+k'} \mathcal{T}^{k k'}_{q}  \,
\end{align}
and
\begin{eqnarray} \label{eq:ColSym1}
\mathcal{T}^{k k'}_{q}
= 
\begin{cases}
{\rm \bf Real }       \,\, \text{ for even } k+k' \\
{\rm \bf Imaginary }  \,\, \text{ for odd } k+k' \text{ and } q \neq 0 \, . \\
{\bf 0}   \text{ for odd }  \,\, k+k' \text{ and } q=0 
\end{cases}
\end{eqnarray}
$\mathcal{R}^{k k'}_{q q'}$ have similar properties  as $\mathcal{T}^{k k'}_{q q'}$.

For simplicity, let us consider a two-level system with unpolarizable ground state $ J_{\alpha_{l}} \!=\! 0$. 
Since the ground state is unpolarizable, we are interested only in atomic polarization of the excited state. Thus, our intention is 
to obtain $\rho^{k}_{q}$ elements describing the state of the excited level which give its atomic  polarization in the reference $\Sigma$ (e.g. Derouich et al. 2007).
We 
focus on the alignment ($\rho^{k}_{q}$ with even $k$)-to-orientation ($\rho^{k}_{q}$ with odd $k$) transfer within the polarizable upper level. Alignment-to-orientation transfer by anisotropic collisions could explain, for instance, solar observations of circular SP   by  L\'opez Ariste et al. (2005).
To demonstrate
the possibility of circular SP creation by collisions, one must determine orientation elements $\rho^{k=odd}_{q}$ to show that they are not equal to zero.
We solve the SEE, given by Eq.~(\ref{eq:SEE1}), in the reference frame $\Sigma$. The Euler angles of the rotation ${\rm R}(\text{-}\gamma_{\rm B},\text{-}\theta_{\rm B},\text{-}\chi_{\rm B})$ are between the magnetic reference  and the frame $\Sigma$.

For the purpose of illustration, we take the total angular momentum of the excited state, $ J_{\alpha_u}$, to be $1$.    Further for simplicity, we take $z_{\rm mag}$ to be in the $\{xz\}_{\rm pert}$-plane, i.e. we set the azimuthal angle $\chi_{\rm B}\!=\!0$.  In what follows,  we   replace  the   notations  $J_{\alpha_{l}} $ and $ J_{\alpha_{u}} $ by 0 and 1, respectively. For example $\mathcal{R}^{11}_1(1,0) $ correspond  to the relaxation rates $\mathcal{R}^{k k'}_{q}$ associated to the loss  of electrons from the level $ J_{\alpha_{u}} \!=\! 1 $ to the level $ J_{\alpha_{l}} \!=\!0$ where $k \!=\! k' \!=\! 1$ and $q \!=\! 1$.
$\mathcal{R}^{11}_1(1) $
are the relaxation rates due to elastic collisions within the same level $J_{\alpha_{i}} \!=\! 1$. Similarly $\mathcal{T}^{20}_0(1,0) $ is the gain of electrons going from the level $ J_{\alpha_{l}} \!=\! 0 $  to the level $ J_{\alpha_{u}} \!=\! 1 $ where $k \!=\! 2$, $k' \!=\! 0$ and $q \!=\! 0$ and $\mathcal{T}^{20}_0(1)$ represents the gain due to electrons transfer from sublevels within the same level. Physically, the collisional relaxation  corresponds to the loss of the atomic $q$-coherence/$k$-order of the level under consideration.  In contrast, the transfer rates corresponds to the gain of coherence or order coming from other levels and sublevels of the same level. 
Collisional contribution to the evolution of the density-matrix elements is due to transfer  and relaxation rates. 
 The full set of SEE  describing the two-level system under consideration  are provided in the Appendix~\ref{app:SEE}. Let us mention that we used the same way to denote the density matrix  elements as for collisional rates, for instance $\rho^1_1(1)$ represents the  density matrix element where $k \!=\! 1$, $q \!=\! 1$ and $ J_{\alpha_{l}} \!=\! 1 $.

Solution of the SEE in the general case, where the three sources of anisotropy discussed above are all present, leads to very large expressions which we do not show here. Instead, we consider some special cases in which two sources of anistropy are present at a time. As we show below, this is enough to illustrate our main point; namely, the breaking of cylindrical symmetry could lead to the emergence of circular SP. 
In addition, 
  we give only the $\rho^{1}_{0}(\alpha_{u} \, J_{\alpha_{u}})$ and $\rho^{1}_{1}(\alpha_{u} \, J_{\alpha_{u}})=$ ${\rm Re}\rho^{1}_{1}(\alpha_{u} \, J_{\alpha_{u}})$ + i ${\rm Im}\rho^{1}_{1}(\alpha_{u} \, J_{\alpha_{u}})$ to  show that it is possible to obtain orientation with $k=1$ (circular polarization) from alignment with $k=2$ (linear polarization). 
Other expressions of 
$\rho^{k}_{q}(\alpha_{i} \, J_{\alpha_{i}})$   can be  obtained from the set of SEE in Appendix~\ref{app:SEE}.

\section{ Solutions of the SEE and  Discussion} \label{sect:discussion}
As discussed above, in a spherically symmetric situation, the atomic polarization, if at all present, can only decrease. Reduction of the symmetry of the problem leads to the formation or increase of SP. For example, linear SP can be created in the presence of anisotropic radiation (e.g. Sect. 10.2 of Landi Degl'Innocenti E. \& Landolfi M. 2004) or anisotropic collisions  (e.g. Sahal-Br\'{e}chot et al. 1996, Vogt et al. 2001). We now show that further reduction in the symmetry of the problem can lead to the generation of circular SP.

\subsection{ Anisotropic collisions \& ainsotropic radiation  field:}
We first consider the case where axially symmetric collisions and an axially symmetric unpolarized radiation field, whose axes of symmetry are in general not parallel to each other, are present.
The SEE describing the situation are obtained from those in Appendix~\ref{app:SEE} by setting $\omega _{\rm L}$=0. These SEE can be easily solved to obtain the density matrix.
The elements of density matrix with $k\!=\!1$ are given by: 
\begin{equation} \label{eq_rho10p}
\rho^1_0(1) = 0 \, ,
\end{equation}
\begin{equation} \label{eq_rho11p}
\rho^1_1(1) =
\text{Re} \rho^1_1(1) \!+\! \text{i} \ \text{Im} \rho^1_1(1)
=
\mathscr{C} B_{01} \big( \text{ImJ}^2_1  \!+\! \text{i} \ \text{ReJ}^2_1 \big) \rho^0_0(0)
= \text{i} \ \mathscr{C} B_{01} \left(\text{J}^2_1\right)^* \rho^0_0(0) \, ,
\end{equation}
where $\mathscr{C}$ is given by 
\begin{equation}   \label{eq_rho11factor}
\mathscr{C} \!\equiv\! - 
\frac{  
\mathcal{C}^{12}_1(1,0)} 
{\sqrt{3} \big[
\big( \mathcal{C}^{11}_1(1,0) \!+\! A_{10} \big)
\big( \mathcal{C}^{22}_1(1,0) \!+\! A_{10} \big)
\!+\!
\mathcal{C}^{12}_1(1,0) \mathcal{C}^{21}_1(1,0) \big]} \, .
\end{equation}
Here $A_{10}$ and $B_{01}$ respectively denote the Einstein coefficients for spontaneous emission and photon absorption characterizing the probability of transitions between the lower level with $J_{\alpha_l}\!=\!0$ and the upper level with $J_{\alpha_u}\!=\!1$.
We have also defined $\mathcal{C}^{kk'}_q(J_{\alpha_i},J_{\alpha_j}) \!\equiv\! \mathcal{R}^{kk'}_q(J_{\alpha_i} \!\!\rightarrow\!\! J_{\alpha_j}) + \mathcal{R}^{kk'}_q(J_{\alpha_i}\!\!\rightarrow\!\! J_{\alpha_i} )  - \mathcal{T}^{kk'}_q(J_{\alpha_i} \!\!\leftarrow\!\! J_{\alpha_i} ) \!\equiv\! \mathcal{R}^{kk'}_q(J_{\alpha_i},J_{\alpha_j}) + \mathcal{R}^{kk'}_q(J_{\alpha_i})  - \mathcal{T}^{kk'}_q(J_{\alpha_i}) $  [e.g. $\mathcal{C}^{12}_1(1,0) \!=\! \mathcal{R}^{12}_1(1,0) + \mathcal{R}^{12}_1(1) - \mathcal{T}^{12}_1(1)$]. 
As can be seen from Eqs.~(\ref{eq_rho11p}) and (\ref{eq_rho11factor}) the $\rho^1_1(1)$ is non-zero, signaling the emergence of circular SP, provided that $\mathcal{C}^{12}_1(1,0)$ and $\rho^0_0(0)$ are different from zero. The rate $\mathcal{C}^{12}_1(1,0)$ is necessarily non-zero given the symmetry conditions explained in Sect.~\ref{sec:theory} (see e.g. Manabe et al. 1979). Futher, the density matrix element of the lower level, $\rho^0_0(0)$, is expected to be different from zero since lifetime of the lower level is usually large compared to the upper level.
This is the case even if $\mathcal{R}^{kk'}_q(J_{\alpha_i})  \!=\! \mathcal{T}^{kk'}_q(J_{\alpha_i})$ as can be verified from the definition of $\mathcal{C}^{kk'}_q(J_{\alpha_i},J_{\alpha_j})$ above.

The generation of circular SP is clearly due to the breaking of cylindrical symmetry of the problem. Had the radiation field been isotropic or having its axis of cylindrical symmetry, $z_{\rm rad}$, parallel/anti-parallel to that of collisions, $z_{\rm pert}$, there would be no coherences in the radiation field $J^2_{q \neq 0} \!=\!0$ (in the frame, $\Sigma$) and hence no emergence of circular SP. Similarly, if the collisions were isotropic, collisional rates with $k \!\neq\! k'$ or with $ q \!\neq\!0$ would vanish. Consequently, there would be no circular SP as can be seen from Eqs.~(\ref{eq_rho11p}) and (\ref{eq_rho11factor}). In  the last two cases, the cylindrical symmetry of the problem is restored and thus there can only be linear SP. In other words, the generation of circular SP is possible only if the whole problem is neither isotropic nor having axial symmetry.

\subsection{  Anisotropic collisions \& oriented magnetic field}
Let us consider another case of broken axial symmetry to further illustrate our point. In this setup we have an ensemble of atoms undergoing axially symmetric collisions in the presence of an oriented magnetic field and isotropic radiation field.
This case is described by the SEE given in Appendix~\ref{app:SEE} while setting $J^2_{q} \!=\!0$. Solution of the SEE in this case holds:
\begin{equation} \label{eq_rho10B}
\rho^1_0(1) \!=\!
\frac{\sqrt{2} g_1 \omega_{\text{L,1}} s_{\theta_{\rm B}} \text{Im} \rho^1_1(1)}
{ [  \mathcal{C}^{11}_0(1,0) + A_{10} ]} \, ,
\end{equation}
\begin{align} \label{eq_rho11ReB}
&\hspace{-1.2cm}\text{Re} \rho^1_1(1) \!=\!\!
\Bigg(\!\!
g_1 \omega_{\text{L,1}} s_{\theta_{\rm B}} \mathcal{C}^{12}_1(1,0)
\Bigg\{\!
g_1^2 \omega_{\text{L,1}}^2
\!\Bigg[\!
\!-\! 4 g_1^2 \omega_{\text{L,1}}^2 c_{\theta_{\rm B}}^4 \big( \mathcal{C}^{11}_0(1,0) \!+\! A_{10} \big) 
\!+\! c_{\theta_{\rm B}}^2
\Big[
2 g_1^2 \omega_{\text{L,1}}^2 s_{\theta_{\rm B}}^2  \big( \mathcal{C}^{11}_0(1,0) \!+\! 2 \mathcal{C}^{22}_1(1,0) \!+\! 3 A_{10} \big) 
\nonumber \\ & 
\hspace{-1.2cm}\!\!+\!
\big( \mathcal{C}^{11}_0(1,0) \!+\! A_{10} \big) 
\!\Big\{\! 
3 A_{10}^2 \!+\! \big[  4  \mathcal{C}^{11}_1(1,0) \!+\! 4 \mathcal{C}^{22}_1(1,0)  \!-\! 2 \mathcal{C}^{22}_2(1,0)  \big] \! A_{10}
\!-\!  [\mathcal{C}^{22}_2(1,0)]^2 \!+\! 4 \mathcal{C}^{12}_1(1,0) \mathcal{C}^{21}_1(1,0) \!+\! 4\mathcal{C}^{11}_1(1,0) \mathcal{C}^{22}_1(1,0) 
\!\Big\}\!
\Big]
\nonumber \\ &
\hspace{-1.2cm}\!\!+\!
s_{\theta_{\rm B}}^2 
\big( \mathcal{C}^{22}_2(1,0)  \!+\! A_{10} \big)
\!\Big\{\!
\big[ \mathcal{C}^{11}_0(1,0) \!+\! \mathcal{C}^{11}_1(1,0) \!+\! \mathcal{C}^{22}_1(1,0) \!+\! \mathcal{C}^{22}_2(1,0)  \big] \! A_{10}
\!+\! 2 A_{10}^2  \!+\! g_1^2 \omega_{\text{L,1}}^2 s_{\theta_{\rm B}}^2 \!+\! \mathcal{C}^{11}_0(1,0) \mathcal{C}^{11}_1(1,0) 
\nonumber \\ &
\hspace{-1.2cm}\!\!+\! \mathcal{C}^{22}_1(1,0) \mathcal{C}^{22}_2(1,0) 
\!\Big\}\!
\Bigg]
\!+\!
\big( \mathcal{C}^{11}_0(1,0)  \!+\! A_{10} \big)
\!\big[ \mathcal{C}^{12}_1(1,0) \mathcal{C}^{21}_1(1,0)
\!+\! \big( \mathcal{C}^{11}_1(1,0)  \!+\! A_{10} \big) \big( \mathcal{C}^{22}_1(1,0)  \!+\! A_{10} \big)
\big] \!
\big( \mathcal{C}^{22}_2(1,0)  \!+\! A_{10} \big)^{\!2}
\!\!\Bigg\}
\nonumber \\ &
\hspace{-1.2cm}\!\!\times\!\!\!
\Bigg\{\!
3 \mathcal{C}^{00}_0(0,1) \big( \mathcal{C}^{00}_0(1,0)  \!+\! A_{10} \big)
\!+\!
B_{01} \text{J}^0_0 \big( 3 \mathcal{C}^{00}_0(1,0)  \!-\! \sqrt{3} \mathcal{T}^{00}_0(0,1) \big)  \!-\! 3 \big( \mathcal{T}^{00}_0(0,1) \!+\! \sqrt{3} A_{10}  \big) \mathcal{T}^{00}_0(1,0)
\!\!\Bigg\}
\!\Bigg) \rho^0_0(0)
%
\Bigg/
%
\Bigg(\!\!
\sqrt{6} \Bigg\{\!\!
4 g_1^6 \omega_{\text{L,1}}^6 c_{\theta_{\rm B}}^6
\nonumber \\ & 
\hspace{-1.2cm}\!\!\times\!
 \! \big( \mathcal{C}^{11}_0(1,0)  \!+\! A_{10} \big)
\!+\!\!
\Big\{\! g_1^2 \omega_{\text{L,1}}^2 s_{\theta_{\rm B}}^2 \! \big(  \mathcal{C}^{11}_1(1,0)  \!+\! A_{10} \big) \!+\! \big[ \mathcal{C}^{12}_1(1,0) \mathcal{C}^{21}_1(1,0)  \!+\! \big( \mathcal{C}^{11}_1(1,0)  \!+\! A_{10} \big) \! \big( \mathcal{C}^{22}_1(1,0)  \!+\! A_{10} \big) \big] \! \big( \mathcal{C}^{22}_2(1,0)  \!+\! A_{10} \big)
\!\Big\}
\nonumber \\ & 
\hspace{-1.2cm}\!\!\times\!
\!\Big\{\!
g_1^4 \omega_{\text{L,1}}^4 s_{\theta_{\rm B}}^4 
\!+\!
g_1^2 \omega_{\text{L,1}}^2 s_{\theta_{\rm B}}^2 \!
\big[  \mathcal{C}^{11}_0(1,0) \mathcal{C}^{11}_1(1,0) \!+\! \mathcal{C}^{22}_1(1,0) \mathcal{C}^{22}_2(1,0) 
\!+\! \big( \mathcal{C}^{11}_0(1,0) \!+\! \mathcal{C}^{11}_1(1,0) \!+\! \mathcal{C}^{22}_1(1,0) \!+\! \mathcal{C}^{22}_2(1,0)  \big) \! A_{10} \!+\!  2 A_{10}^2  \big] 
\nonumber \\ & 
\hspace{-1.2cm}\!\!+\! 
\big( \mathcal{C}^{11}_0(1,0)  \!+\! A_{10} \big) \big( \mathcal{C}^{11}_0(1,0)  \!+\! A_{10} \big) 
\! \big[ \mathcal{C}^{12}_1(1,0) \mathcal{C}^{21}_1(1,0)  \!+\! \big( \mathcal{C}^{11}_1(1,0)  \!+\! A_{10} \big) \! \big( \mathcal{C}^{22}_1(1,0)  \!+\! A_{10} \big) \big] \! 
\big( \mathcal{C}^{22}_2(1,0)  \!+\! A_{10} \big)
\!\!\Big\}\!
\nonumber \\ & 
\hspace{-1.2cm}\!\!+\!
c_{\theta_{\rm B}}^4 
\!\!\Big\{\!
4 g_1^6 s_{\theta_{\rm B}}^2 \omega_{\text{L,1}}^6 \! \big( \mathcal{C}^{11}_1(1,0) \!-\! \mathcal{C}^{11}_0(1,0)  \big)
\!+\! 
g_1^4 \omega_{\text{L,1}}^4 \! \big(  \mathcal{C}^{11}_0(1,0)  \!+\! A_{10} \big) \!
\Big[\! 9 A_{10}^2 \!+\!
2 \big[ 4 \big(  \mathcal{C}^{11}_1(1,0) \!+\! \mathcal{C}^{22}_1(1,0)  \big)  \!+\! \mathcal{C}^{22}_2(1,0)  \big] \! A_{10} 
\nonumber \\ & 
\hspace{-1.2cm}\!\!+\! 
[\mathcal{C}^{22}_2(1,0)  ]^2 \!\!+\! 4 \big( [\mathcal{C}^{11}_1(1,0)]^2 \!\!+\! [\mathcal{C}^{22}_1(1,0)]^2 \!\!-\! 2 \mathcal{C}^{12}_1(1,0) \mathcal{C}^{21}_1(1,0) \big)  \!\Big] \!\Big\}
\!+\!
g_1^6 c_{\theta_{\rm B}}^2
\!\Big\{\!
\omega_{\text{L,1}}^6 s_{\theta_{\rm B}}^4 
\! \big(  \mathcal{C}^{11}_0(1,0) \!-\! 4 \mathcal{C}^{11}_1(1,0)  \!-\! 3A_{10} \big)
\nonumber \\ &
\hspace{-1.2cm}\!\!+\!
g_1^4 s_{\theta_{\rm B}}^2 \! \omega_{\text{L,1}}^4 \!
\Big[
3 A_{10}^3 \!-\!
\big(  2 \mathcal{C}^{11}_0(1,0) \!+\! 3 \mathcal{C}^{11}_1(1,0) \!-\! 10 \mathcal{C}^{22}_1(1,0) \!-\! 4 \mathcal{C}^{22}_2(1,0)  \big) \! A_{10}^2 \!+\!
\big\{\! 4 [\mathcal{C}^{22}_1(1,0)]^2 \!\!+\! 2 \big( \mathcal{C}^{11}_0(1,0) \!+\! 4 \mathcal{C}^{11}_1(1,0) \big) \mathcal{C}^{22}_1(1,0) 
\nonumber \\ &
\hspace{-1.2cm}\!\!+\! 
[\mathcal{C}^{22}_2(1,0)]^2  \!\!-\! 4 \big( 2 \mathcal{C}^{11}_0(1,0) \!+\! \mathcal{C}^{11}_1(1,0) \big) \mathcal{C}^{11}_1(1,0) \!+\! 8 \mathcal{C}^{12}_1(1,0) \mathcal{C}^{21}_1(1,0)
\!+\!
2 \big( \mathcal{C}^{11}_0(1,0) \!+\! \mathcal{C}^{11}_1(1,0) \!+\! \mathcal{C}^{22}_1(1,0) \big) \mathcal{C}^{22}_2(1,0) 
\!\big\}\!  A_{10}
\nonumber \\ &
\hspace{-1.2cm}\!\!+\!
 \mathcal{C}^{11}_1(1,0) [\mathcal{C}^{22}_2(1,0)]^2 \!\!+\!
4 \mathcal{C}^{22}_1(1,0)
\big( \mathcal{C}^{12}_1(1,0) \mathcal{C}^{21}_1(1,0) \!+\! \mathcal{C}^{11}_1(1,0) \mathcal{C}^{22}_1(1,0) \big) \!+\!
\mathcal{C}^{11}_0(1,0) \!
\big\{\! \!-\! 4 [\mathcal{C}^{11}_1(1,0)]^2 \!\!+\! 4 \mathcal{C}^{12}_1(1,0) \mathcal{C}^{21}_1(1,0)  
\nonumber \\ &
\hspace{-1.2cm}\!\!+\! 
  2 \mathcal{C}^{22}_1(1,0) \mathcal{C}^{22}_2(1,0)  \!\big\}
\!\Big]\! \!+\!
g_1^2 \omega_{\text{L,1}}^2 \! \big(   \mathcal{C}^{11}_0(1,0)  \!+\! A_{10} \big) 
\! \Big[\! 6 A_{10}^4 \!+\!
 2 \big\{\! 5 \big( \mathcal{C}^{11}_1(1,0) \!+\! \mathcal{C}^{22}_1(1,0) \big) \!+\! 2 \mathcal{C}^{22}_2(1,0) 
\!\big\} \!  A_{10}^3 \!+\!
\big\{\!   5 [\mathcal{C}^{11}_1(1,0)]^2  
\nonumber \\ &
\hspace{-1.2cm}\!\!+\! 16 \mathcal{C}^{22}_1(1,0) \mathcal{C}^{11}_1(1,0) \!+\!  5 [\mathcal{C}^{22}_1(1,0)]^2 \!\!+\! 2 [\mathcal{C}^{22}_2(1,0)]^2 \!\!+\! 6 \mathcal{C}^{12}_1(1,0) \mathcal{C}^{21}_1(1,0)  \!+\! 4 \big( \mathcal{C}^{11}_1(1,0) \!+\! \mathcal{C}^{22}_1(1,0) \big)  \mathcal{C}^{22}_2(1,0)  \!\big\} \! A_{10}^2
\nonumber \\ &
\hspace{-1.2cm}\!\!+\!
2 \big\{\! \big( \mathcal{C}^{11}_1(1,0) \!+\! \mathcal{C}^{22}_1(1,0) \big) \! [\mathcal{C}^{22}_2(1,0)]^2
\!+\! \big( [\mathcal{C}^{11}_1(1,0)]^2 \!+\! [\mathcal{C}^{22}_1(1,0)]^2 \!-\! 2 \mathcal{C}^{12}_1(1,0) \mathcal{C}^{21}_1(1,0) \big) \mathcal{C}^{22}_2(1,0) \!+\! 4 \big( \mathcal{C}^{11}_1(1,0) \!+\! \mathcal{C}^{22}_1(1,0) \big) 
\nonumber \\ &
\hspace{-1.2cm}\!\!\times\!\!
\big( \mathcal{C}^{12}_1(1,0) \mathcal{C}^{21}_1(1,0) \!+\! \mathcal{C}^{11}_1(1,0) \mathcal{C}^{22}_1(1,0) \big)  
\!\big\} \!
A_{10}
\!+\!
\big( [\mathcal{C}^{11}_1(1,0)^2 \!+\! [\mathcal{C}^{22}_1(1,0)]^2 \!-\! 2 \mathcal{C}^{12}_1(1,0) \mathcal{C}^{21}_1(1,0) \big) [\mathcal{C}^{22}_2(1,0)]^2 
\nonumber \\ &
\hspace{-1.2cm}\!\!+\!
 4 \big( \mathcal{C}^{12}_1(1,0) \mathcal{C}^{21}_1(1,0) \!+\! \mathcal{C}^{11}_1(1,0) \mathcal{C}^{22}_1(1,0) \big)^{\!2} 
\!\Big] 
\!\Big\}
\!\!\Bigg\}
\Bigg\{\!\! \mathcal{C}^{02}_0(1,0)  \mathcal{T}^{00}_0(0,1) \!+\! \big( \! \sqrt{3} \mathcal{C}^{02}_0(1,0)  \!-\! \mathcal{T}^{02}_0(0,1) \big) \! A_{10} \!-\! \mathcal{C}^{00}_0(1,0)  \mathcal{T}^{02}_0(0,1)
\Bigg\}
\!\!\Bigg) \, ,
\end{align}
\begin{align} \label{eq_rho11ImB}
&\hspace{-1.2cm}\text{Im} \rho^1_1(1) \!=\! -
\Bigg(\!\!
\sqrt{\frac{2}{3}} 
g_1^2 \omega_{\text{L,1}}^2 s_{\theta_{\rm B}} c_{\theta_{\rm B}}
\mathcal{C}^{12}_1(1,0) 
\Big[
g_1^2 \omega_{\text{L,1}}^2
\big\{ 4 c_{\theta_{\rm B}}^2  \big( \mathcal{C}^{11}_1(1,0) \!+\! \mathcal{C}^{22}_1(1,0)  \!+\! 2 A_{10} \big) 
\!+\! s_{\theta_{\rm B}}^2 \big( \!-\! 2 \mathcal{C}^{11}_1(1,0) \!+\! \mathcal{C}^{22}_2(1,0)  \!-\! A_{10} \big) \big\}
\nonumber \\ &
\hspace{-1.2cm}\!\!+\!\!
\big( \mathcal{C}^{11}_1(1,0) \!+\! \mathcal{C}^{22}_1(1,0)  \!+\! 2 A_{10} \big) \big( \mathcal{C}^{22}_2(1,0)  \!+\! A_{10} \big)^2
\Big]
\Big[
\!-\! 3 \mathcal{C}^{00}_0(0,1)  \big( \mathcal{C}^{00}_0(1,0)  \!+\! A_{10} \big)
\!+\! B_{01} \text{J}^0_0 
\big( \sqrt{3} \mathcal{T}^{00}_0(0,1) \!-\! 3 \mathcal{C}^{00}_0(1,0)  \big)
\nonumber \\ &
\hspace{-1.2cm}\!\!+\!
3 \big( \mathcal{T}^{00}_0(0,1) \!+\! \sqrt{3} A_{10} \big) \mathcal{T}^{00}_0(1,0)
\Big]
\!\!\Bigg)
\rho ^0_0(0)
%
\Bigg/
%
\Bigg(\!\!
\Bigg\{\!\!
\Big[
\!-\! g_1^2 \omega_{\text{L,1}}^2 c_{\theta_{\rm B}}^2 
\big( 2 \mathcal{C}^{22}_1(1,0) \!+\! \mathcal{C}^{22}_2(1,0)  \!+\! 3 A_{10} \big) 
\!+\!
\mathcal{C}^{12}_1(1,0) \mathcal{C}^{21}_1(1,0) 
\big( \mathcal{C}^{22}_2(1,0)  \!+\! A_{10} \big)
\nonumber \\ &
\hspace{-1.2cm}\!\!+\!\!
\big\{
g_1^2 \omega_{\text{L,1}}^2 s_{\theta_{\rm B}}^2 \!+\! \big( \mathcal{C}^{11}_0(1,0)  \!+\! A_{10} \big) \big( \mathcal{C}^{11}_1(1,0)  \!+\! A_{10} \big)
\big\}
\big\{
g_1^2 \omega_{\text{L,1}}^2 \big( 1 \!-\! 3 c_{\theta_{\rm B}}^2 \big)
\!+\! \big( \mathcal{C}^{22}_1(1,0)  \!+\! A_{10} \big) \big( \mathcal{C}^{22}_2(1,0)  \!+\! A_{10} \big)
\big\}      \big/    \big\{ \mathcal{C}^{11}_0(1,0) \!+\! A_{10} \big\}
\Big]
\nonumber \\ &
\hspace{-1.2cm}\!\!\times\!\!
\Big[
g_1^2 \omega_{\text{L,1}}^2
\big\{ 4 c_{\theta_{\rm B}}^2 \big( \mathcal{C}^{11}_1(1,0) \!+\! \mathcal{C}^{22}_1(1,0)  \!+\! 2 A_{10} \big)
\!+\!
s_{\theta_{\rm B}}^2  \big( \!-\! 2\mathcal{C}^{11}_1(1,0) \!+\! \mathcal{C}^{22}_2(1,0)  \!-\! A_{10} \big) \big\}
\!+\!
\big( \mathcal{C}^{11}_1(1,0) \!+\! \mathcal{C}^{22}_1(1,0)  \!+\! 2 A_{10} \big) 
\nonumber \\ &
\hspace{-1.2cm}\!\!\times\!\!
\big( \mathcal{C}^{22}_2(1,0)  \!+\! A_{10} \! \big)^{\!2}
\!\Big] 
\!\!+\!\!
\Big[ g_1^2 \omega_{\text{L,1}}^2 \big( 3 c_{\theta_{\rm B}}^2 \!-\! 1 \big)
\!-\! 2 \mathcal{C}^{12}_1(1,0) \mathcal{C}^{21}_1(1,0)
\!-\! 2 \mathcal{C}^{11}_1(1,0) \mathcal{C}^{22}_1(1,0) 
\!-\! \big\{\! 3 \big(  \mathcal{C}^{11}_1(1,0)
\!+\! \mathcal{C}^{22}_1(1,0) \big) \!+\! 2 \mathcal{C}^{22}_2(1,0)  \!\big\} \! A_{10}
\nonumber \\ &
\hspace{-1.2cm}\!\!-\!
\big( \mathcal{C}^{11}_1(1,0) \!+\! \mathcal{C}^{22}_1(1,0) \big) \mathcal{C}^{22}_2(1,0)  \!-\! 4 A_{10}^2
\!\Big] 
\!
\Big[ 
g_1^2 \omega_{\text{L,1}}^2 c_{\theta_{\rm B}}^2  
\big\{[ 2 g_1^2 \omega_{\text{L,1}}^2 \big(1 \!-\! 3 c_{\theta_{\rm B}}^2\big) \!-\! \big( \mathcal{C}^{22}_2(1,0) \!+\! A_{10} \big)^{\!2} \big\}
\!+\! \mathcal{C}^{12}_1(1,0) \mathcal{C}^{21}_1(1,0) 
\nonumber \\ &
\hspace{-1.2cm}\!\!\times\!\!
\big\{ 4  g_1^2 \omega_{\text{L,1}}^2 c_{\theta_{\rm B}}^2 \!+\! \big( \mathcal{C}^{22}_2(1,0)  \!+\! A_{10} \big)^{\!2} \big\}
\!+\!\!
\big\{
g_1^2 \omega _{\text{L,1}}^2 s_{\theta_{\rm B}}^2
\!+\!
\big( \mathcal{C}^{11}_0(1,0)  \!+\! A_{10} \big)
\big( \mathcal{C}^{11}_1(1,0)  \!+\! A_{10} \big)
\big\}
\big\{
g_1^2 \omega_{\text{L,1}}^2
\big[ 4 c_{\theta_{\rm B}}^2 \big( \mathcal{C}^{22}_1(1,0)  \!+\! A_{10} \big)
\nonumber \\ &
\hspace{-1.2cm}\!\!+\!\! 
s_{\theta_{\rm B}}^2 ( \mathcal{C}^{22}_2(1,0)  \!+\! A_{10} ) \big]
\!+\!\!
\big(  \mathcal{C}^{22}_1(1,0)  \!+\! A_{10} \big) \big( \mathcal{C}^{22}_2(1,0)  \!+\! A_{10} \big)^{\!2}
\big\}    \big/  \big\{ \mathcal{C}^{11}_0(1,0)  \!+\! A_{10} \big\}
\Big]
\!\!\Bigg\}
\nonumber \\ &
\hspace{-1.2cm}\!\!\times\!\!
\!
\Bigg\{\!\! \mathcal{C}^{02}_0(1,0)  \mathcal{T}^{00}_0(0,1) \!+\! A_{10}  \big( \sqrt{3} \mathcal{C}^{02}_0(1,0)  \!-\! \mathcal{T}^{02}_0(0,1) \big) \!-\!  \mathcal{C}^{00}_0(1,0)  \mathcal{T}^{02}_0(0,1)
\!\!\Bigg\}
\!\!\Bigg) \,,
\end{align}
where we have defined $s_{\theta_{\rm B}} \!\equiv\! \sin{\theta_{\rm B}}$ and $c_{\theta_{\rm B}} \!\equiv\! \cos{\theta_{\rm B}}$. In the case at hand, circular SP is generated which is again attributed to the breaking of axial symmetry of the problem. This can be   verified from Eqs.~(\ref{eq_rho10B}), (\ref{eq_rho11ReB}) and (\ref{eq_rho11ImB}) by setting $s_{\theta_{\rm B}} \!=\!0$ or $\pi$, i.e. by making $z_{\rm pert}$ and $z_{\rm mag}$ respectively parallel or anti-parallel, which  yields
$$ \rho^1_0(1) \!=\! \text{Re}\rho^1_1(1) \!=\! \text{Im}\rho^1_1(1) \!=\! 0 \,. $$
In other words, restoring the cylindrical symmetry of the problem results in a vanishing circular SP. It can be  verified from Eqs.~(\ref{eq_rho10B}), (\ref{eq_rho11ReB}) and (\ref{eq_rho11ImB}) that the circular SP would be present if $\theta_{\rm B}$ is neither zero nor $\pi$. In particular, we have  verified that for the special case of $\theta_{\rm B} \!=\! \pi/2$, that $\rho^1_0(1) \!=\! \text{Im}\rho^1_1(1) \!=\! 0$  but $\text{Re} \rho^1_1(1) \!\neq\!0$. 

\subsection{  Anisotropic radiation \& oriented magnetic field }
  For completeness, let us also consider the case in which an ensemble of atoms is illuminated by an anisotropic radiation in the presence of  an oriented magnetic field. We could also allow the atoms to undergo isotropic collisions. The SEE describing this situation is obtained from those in Appendix~\ref{app:SEE} by setting all collisional rates with $k \!\neq\! k'$ or with $q \!\neq\!0$ to zero. Solving for the density matrix elements, one can  verify that
$$ \rho^1_0(1) \!=\! \text{Re} \rho^1_1 \!=\! \text{Im} \rho^1_1 \!=\! 0 \,.$$
Clearly, the generation of circular SP is not possible in this case despite the breaking of the axial symmetry of the problem. This is due the fact that a weak magnetic field cannot cause the mixing of density matrix elements with different order, $k$, besides
the restriction imposed by selection rules on the possible optical transitions which prevents the mixing between odd- and even-order density matrix elements ( see e.g. Sects. 7.11 and 10.8 Landi Degl'Innocenti E. \& Landolfi M. 2004). The later obstacle is not present in the case of anisotropic collisions. That is the reason, in the case of anisotropic collisions, the breaking of cylindrical symmetry leads to the generation of circular SP  whereas there is no creation of circular SP if the anisotropy in collisions is replaced by a deterministic weak magnetic field.

\section{Conclusions} \label{sect:conclusion} 
We formulated circularly polarizing effect of anisotropic  collisions
in the presence of   anisotropic radiation field and/or  deterministic magnetic field.  
In particular, we   show  the possibility of
creation of atomic circular SP if the density of
perturbers is sufficient
for anisotropic collisions to be effective.  
This physical situation can
occur in a plasma where
charged particles (e.g. protons or electrons)  
move in a direction  
different from   that photons most frequently are moving in and/or different from 
that of the magnetic field, in a way that cylindrical symmetry of the problem is broken.

In order to contribute to interpretations of  chromospheric H$ \alpha $ line observations
of hydrogen  (L\'opez Ariste et al.  2005, Ramelli et al.  2005), it is important to  calculate the relaxation and transfer rates  due to anisotropic collisions of hydrogen atoms with
  electrons. Then,
it is necessary to introduce them in a code of resolution of the SEE in order to determine the circular polarization. In addition, 
in the low corona, 
it is now well-established that the  velocity distributions of the solar wind's electrons, protons  and heavy ions are  non-thermal, meaning that they are anisotropic and cannot
be described by a Maxwellian distribution  (e.g.   
 Pilipp et al. 1987 and Pierrard et al. 2001).   Different   models are proposed to represent those distributions, like  
bi-Maxwellian  or  kappa distributions   (e.g. Maksimovic et al. 1997).  Solar wind diagnostics are traditionally based on spectroscopic analysis, which uses only 
the Stokes-$I$ measurements.  Our results can be used to gain better understanding on the solar wind physics since the polarization is very sensitive to the anisotropic part of velocity distributions.

\begin{acknowledgements}
 We thank the anonymous reviewer for his invaluable comments which helped us improve the presentation 
 of this work. This project was funded by the Deanship of Scientific Research (DSR) at King Abdulaziz University, Jeddah, under grant no. (G:348-130-1440). The authors, therefore, acknowledge with thanks DSR for technical and financial support.
\end{acknowledgements}

\appendix				

\section{SEE} \label{app:SEE}

\noindent
Assuming steady state and making use of conjugation properties of the density matrix, $\left[\rho^k_{q} (\alpha_i J_{\alpha_i})\right]^{*} \!=\! (\text{-}1)^{q} \rho^k_{\text{-}q}(\alpha_i J_{\alpha_i}) $,   and the radiation  field tensor, $\left[{\rm J}^{k_{\rm r}}_{q_{\rm r}} (\nu_{\alpha_{i} J_{\alpha_{i}},jJ_{\alpha_{j}} }) \right]^{*} \!=\! (\text{-}1)^{q_{\rm r}} {\rm J}^{k_{\rm r}}_{\text{-}q_{\rm r}} (\nu_{\alpha_{i} J_{\alpha_{i}},jJ_{\alpha_{j}} }) $, and  the symmetry properties of the collision rates,  given by Eqs.~(\ref{eq:ColSym1}) and (\ref{eq:ColSym2}),
the set of   coupled  SEE can be written as (where $J_{\alpha_{i}} =0$ and $J_{\alpha_{j}}=1$)\footnote{Note that due to symmetry there are some redundant equations. For brevity, we do not show these equations here.}:
\begin{eqnarray}
&&\!\!\!\!
\big(  \mathcal{C}^{00}_0(0,1) \!+\!  B_{01} {\rm J}^0_0 
\big) \rho^0_0(0)
\!-\!
\big( \mathcal{T}^{00}_0(0,1) \!+\! \sqrt{3}  A_{10}  \big) \rho^0_0(1)
\!-\!
\mathcal{T}^{02}_0(0,1) \rho^2_0(1)
\!=\! 0
\\&&\!\!\!\!
\big( \mathcal{T}^{00}_0(1,0) \!+\! \frac{1}{\sqrt{3}}  B_{01} {\rm J}^0_0 \big) \rho^0_0(0) 
\!-\!
\big( \mathcal{C}^{00}_0(1,0) \!+\!  A_{10} \big) \rho^0_0(1)
\!-\!
\mathcal{C}^{02}_0(1,0) \rho^2_0(1)
\!=\! 0
\\&&\!\!\!\!
\sqrt{2} g_1 \omega_{{\rm L},1} s_{\theta_{\rm B}} \text{Im}\rho^1_1(1)
\!-\!
\big( \mathcal{C}^{11}_0(1,0) \!+\! A_{10} \big) \rho^1_0(1)
\!=\! 0
\\&&\!\!\!\!
g_1 \omega_{{\rm L},1} c_{\theta_{\rm B}}  \text{Im}\rho^1_1(1)
\!-\!
\big( \mathcal{C}^{11}_1(1,0) \!+\! A_{10} \big) \text{Re}\rho^1_1(1)
\!+\!
\mathcal{C}^{12}_1(1,0) \text{Im}\rho^2_1(1)
\!=\! 0
\\&&\!\!\!\!
g_1 \omega_{{\rm L},1} \big( s_{\theta_{\rm B}} \text{Im}\rho^2_1(1) \!+\! 2 c_{\theta_{\rm B}}  \text{Im}\rho^2_2(1) \big)
\!-\!
\big(  \mathcal{C}^{22}_2(1,0) \!+\!  A_{10} \big) \text{Re}\rho^2_2(1)
\!+\!
\frac{1}{\sqrt{3}} B_{01} {\rm Re} {\rm J}^2_2 \rho^0_0(0)
\!=\! 0
\\&&\!\!\!\!
\!g_1 \omega_{{\rm L},1} \big( c_{\theta_{\rm B}} \text{Im}\rho^2_1(1) \!+\! s_{\theta_{\rm B}} \text{Im}\rho^2_2(1) \big)
\!+\!
\mathcal{C}^{21}_1(1,0) \text{Im}\rho^1_1(1)
\!-\!
\big( 
\mathcal{C}^{22}_1(1,0) \!+\! A_{10} \big) \text{Re}\rho^2_1(1)
\nonumber\\&&\!\!\!\!
\!+
\frac{1}{\sqrt{3}} B_{01} {\rm Re} {\rm J}^2_1 \rho^0_0(0)
\!=\! 0
\\&&\!\!\!\!
\sqrt{6} g_1 \omega_{{\rm L},1} \!  s_{\theta_{\rm B}} \text{Im}\rho^2_1(1)
\!+\! 
\big( \mathcal{T}^{20}_0(1,0) \!+\! \frac{1}{\sqrt{3}}  B_{01} {\rm J}^2_0  \big) \rho^0_0(0) 
\!-\!
\mathcal{C}^{20}_0(1,0) \rho^0_0(1)
\nonumber\\&&\!\!\!\!
\!
\!-\!
\big(  \mathcal{C}^{22}_0(1,0) \!+\!  A_{10} \big) \rho^2_0(1)
\!=\! 0
\\&&\!\!\!\!
\frac{1}{2} g_1 \omega_{{\rm L},1} \big( \sqrt{2} s_{\theta_{\rm B}} \rho^1_0(1) \!+\! 2 c_{\theta_{\rm B}} \text{Re}\rho^1_1(1) \big)
\!+\!
\big( \mathcal{C}^{11}_1(1,0) \!+\! A_{10} \big) \text{Im}\rho^1_1(1)
\!+\! \mathcal{C}^{12}_1(1,0) \text{Re}\rho^2_1(1)
\!=\! 0
\\&&\!\!\!\!
g_1 \omega_{{\rm L},1} \big( s_{\theta_{\rm B}} \text{Re}\rho^2_1(1) \!+\! 2 c_{\theta_{\rm B}} \text{Re}\rho^2_2(1) \big)
\!+\!
\big( \mathcal{C}^{22}_2(1,0) \!+\! A_{10} \big) \text{Im}\rho^2_2(1)
\!+\!
\frac{1}{\sqrt{3}} B_{01} {\rm Im} {\rm J}^2_2 \rho^0_0(0)
\!=\! 0
\\&&\!\!\!\!
\frac{1}{2} g_1 \omega_{{\rm L},1}\! 
\Big[ s_{\theta_{\rm B}} \big( \sqrt{6} \rho^2_0(1) \!+\! 2 \text{Re}\rho^2_2(1) \big)
\!+\! 2 c_{\theta_{\rm B}} \text{Re}\rho^2_1(1) \Big]
\!+\!
\mathcal{C}^{21}_1(1,0) \text{Re}\rho^1_1(1)
\!+\!
\big( \mathcal{C}^{22}_1(1,0) \!+\! A_{10} \big) \text{Im}\rho^2_1(1)
\nonumber\\&&\!\!\!\!
\!+
\frac{1}{\sqrt{3}} B_{01} {\rm Im} {\rm J}^2_1 \rho^0_0(0)
\!=\! 0
\end{eqnarray}
where we have defined  $\mathcal{C}^{kk'}_q(J_{\alpha_i},J_{\alpha_j}) \!\equiv\! \mathcal{R}^{kk'}_q(J_{\alpha_i} \!\!\rightarrow\!\! J_{\alpha_j}) \!+\! \mathcal{R}^{kk'}_q(J_{\alpha_i}\!\!\rightarrow\!\! J_{\alpha_i} )  \!-\! \mathcal{T}^{kk'}_q(J_{\alpha_i} \!\!\leftarrow\!\! J_{\alpha_i} ) \!\equiv\! \mathcal{R}^{kk'}_q(J_{\alpha_i},J_{\alpha_j}) \!+\! \mathcal{R}^{kk'}_q(J_{\alpha_i})  \!-\! \mathcal{T}^{kk'}_q(J_{\alpha_i}) $  [e.g. $\mathcal{C}^{12}_1(1,0) \!=\! \mathcal{R}^{12}_1(1,0) \!+\! \mathcal{R}^{12}_1(1) \!-\! \mathcal{T}^{12}_1(1)$], $s_{\theta_{\rm B}} \!\equiv\! \sin\!\theta_{\rm B}$ and $c_{\theta_{\rm B}} \!\equiv\! \cos\!\theta_{\rm B}$. The system of equations above, having a zero determinant, is not closed; consequently, one cannot solve for all density matrix elements. 
To overcome this issue, the usual practice is to add the trace equation, i.e $\sum_{i} \sqrt{2 J_{\alpha_i} \!+\! 1} \rho^0_0(J_{\alpha_i}) \!=\! N$ with $N$ being the population number, to the system of equations in order to enable the solution for all density matrix elements. 
However, in the case at hand we are interested only in the orientation terms, $\rho^{k=1}_q$. Therefore, we solve the SEE to obtain $\rho^{k=1}_q$ in terms of the population of the lower level, $\rho^0_0(0)$, which is expected to be non-zero since the lifetime of the lower level is large compared to the upper level. For this purpose we use the algebraic program Mathematica.


\label{lastpage}


\begin{thebibliography}{99}




\bibitem[Bommier (1978)]{Bommier+1978} Bommier V.\& Sahal-Br\'{e}chot S., 1978, A\&A, 69, 57.

 
\bibitem[Aleman (2018)]{Aleman+2018} del Pino Alem\'an T., Trujillo Bueno J., Step\'an J. \& Shchukina N., 2018, ApJ, 863, 20.


\bibitem[Derouich (2003)]{Derouich+2003}   Derouich  M., Sahal-Br\'echot  S., Barklem  P. S.,  O\~{O}Mara  B. J., 2003, A\&A, 404, 763 

\bibitem[Derouich (2007)]{Derouich+2007}
Derouich M., 2007, A\&A, 466, 683. 

\bibitem[Derouich et al (2007)]{Derouich-et-al+2007} Derouich M.,Trujillo Bueno J. \& Manso Sainz R., 2007, A\&A 472, 269-275.

\bibitem[D'yakonov (1978)]{Dyakonov+Perel+1978} D'yakonov M. I. \& Perel V. I., 1978, Sixth Internat. Conf. Atomic phys., Proceedings, Riga, 410.


\bibitem[Hanle (1924)]{Hanle+1924} Hanle W., 1924, Z. Phys., 30, 93  


\bibitem[Landi Degl'Innocenti (2004)]{DeglInnocenti+2004} Landi Degl'Innocenti E. \& Landolfi M., 2004, Polarization in Spectral Lines, Astrophysics and Space Science Library, Vol. 307.


\bibitem[Lopez (2005)]{Lopez_05} L\'{o}pez Ariste A., Casini R., Paletou F., Tomczyk S., Lites B. W., Semel M., Landi Degl'Innocenti E., Trujillo Bueno J., Balasubramaniam K. S., 2005, ApJ,   621, L145.


\bibitem[Manabe (1979)]{Manabe+etal+1979} Manabe T., Yabuzaki T. \& Ogawa T., 1979, Phys. Rev. A, 20, 5, 1946.



\bibitem[Manabe (1981)]{Manabe_81} Manabe T.,  Yabuzaki T., \&  Ogawa T., 1981, Phys. Rev. Lett., 46, 637.   


\bibitem[Maksimovic (1997)]{Maksimovic_1997} Maksimovic M., Pierrard V. \& Lemaire J. F.,  1997,
A\&A,  324, 725-734.


\bibitem[Pierrard (2001)]{Pierrard_01}    Pierrard V. \& Lamy H., 2001, 
Sol. Phys., vol. 216, 47-58.


\bibitem[Pilipp (1987)]{Pilipp_1987}    Pilipp W. G., Muehlhaeuser K.-H., Miggenrieder H., Montgomery M. D. \& Rosenbauer H. , 1987,
J. Geophys. Res., vol. 92,  1075-1092.

\bibitem[Omont (1977)]{Omont+1977} Omont A., 1977, Prog. Quantum Electron., 5, 69.


\bibitem[Ramelli (2005)]{Ramelli_05} Ramelli R., Bianda M., Trujillo Bueno J., Merenda L. \& Stenflo J. O., 2005, Proc. International Scientific Conference on Chromospheric and Coronal Magnetic Fields (ESA SP-596).  Innes D. E., Lagg  A. \&  Solanki S. K. (eds), Published on CDROM, p.82.1.


\bibitem[Petrashen' (1993)]{Petrashen_93} Petrashen A. G., Rebane V. N. \& Rebane T. K., 1993, JETP, 77, 2, 187. 




\bibitem[Trujillo Bueno (2002)]{Bueno+202} Trujillo Bueno J., 2002, 
arXiv:astro-ph/0202328.




\bibitem[Sahal-Br\'{e}chot (1977)]{Sahal+1977} Sahal-Br\'{e}chot S., 1977, ApJ., 213, 887. 

\bibitem[Sahal-Br\'{e}chot (1996)]{Sahal+1996} Sahal-Br\'{e}chot S., Vogt E., Thoraval S. \& Diedhiou I., 1996, A\&A, 309, 317-334.


\bibitem[Vogt (2001)]{Vogt_2001} Vogt E., Sahal-Br\'{e}chot S. \& Bommier V., 2001, A\&A, 374, 1127-1134.


%
\end{thebibliography}
\end{document}